# An Interpretable Latent Space reveals changing dynamics of European heatwaves

Tamara Happé[1*], Jasper S. Wijnands[2], Paolo Scussolini[1], Peter Pfleiderer[3] and Dim Coumou[1]

[1] Institute for Environmental Studies, Vrije Universiteit Amsterdam, Amsterdam, the Netherlands
[2] Royal Netherlands Meteorological Institute (KNMI), de Bilt, the Netherlands
[3] Leipzig Institute for Meteorology (LIM), Universität Leipzig, Leipzig, Germany

*Author to whom any correspondence should be addressed.

E-mail: t.happe@vu.nl

**Abstract**

Due to climate change, heatwaves are becoming more frequent and intense, with western Europe experiencing the strongest trends in the Northern Hemisphere mid-latitudes. Part of the temperature trends are caused by circulation changes, which are not accurately captured in climate models. Here we deploy Deep Learning techniques to classify European heatwaves based on their atmospheric circulation and to study their associated changes over time. We use a Variational Autoencoder (VAE) to reduce the dimensionality of the heatwave samples, after which we cluster them on their extraced features. The VAE is trained on large ensemble climate model simulations and we show that the VAE generalizes well to observed heatwave circulations in ERA5 reanalysis, without the need fLeor transfer learning. The circulation features relevant for heatwaves in ERA5 are consistent with the climate model heatwaves. Regression analysis reveals that the Atlantic Plume type of heatwaves are becoming more frequent over time, while the Atlantic High heatwaves are becoming less frequent. We introduce new and straightforward interpretability methods to study the latent space, including feature importance identification and changes over time. We investigate which circulation features are associated with the most important nodes in the latent space and how the latent space changes over time. For example, we find that the Atlantic Low heatwave shows a deepening of the low pressure system off the Atlantic coast over time. Each heatwave type is undergoing unique changes in their circulation, highlighting the necessity to study each heatwave type separately. Our method can furthermore be used to boost specific aspects of extreme events, and we illustrate how heatwave circulation could change in the future if the current trends persist, with in some cases an intensification of features.

## 1. Introduction

Climate change is altering the frequency and intensity of heatwaves around the world, with western European heat extremes rising faster than elsewhere (Vautard et al., 2023). Heat extremes cause a variety of different impacts: agricultural losses, energy grid disruptions, ecosystem damages, and human health issues and mortality increase (Barriopedro, et al. 2023). For example, the early July 2025 heatwave in western Europe, contributed to increased ozone air pollution and increased risk of wildfires (Copernicus, 2025). During this heatwave, surface temperatures exceeded 40°C in multiple countries, even reaching 46°C in

Spain and Portugal. This specific event was linked to a persistent high-pressure system, with warm air flowing northward from the Saharan region. Such dynamics have been shown to increase since 1950, leading to additional 1.3°C temperature increase in western Europe per degree of global warming (Vautard et al., 2023).

While the thermodynamic contribution to temperature increase is well understood (IPCC - Arias et al., 2021), the contribution of changes in atmospheric dynamics to the rising temperature is generally less known (Shaw et al., 2024). At least part of the increase in western European heat extremes has been attributed to changes in atmospheric dynamics (Singh et al., 2023; Vautard et al., 2023), such as a double jet configuration (Rousi et al., 2022). Concerningly, it was also found that state-of-the-art climate model simulations do not show the same changes in atmospheric circulation type with global warming, highlighting a failure of climate models to capture this observed phenomenon (Vautard et al. 2023, Happé et al., 2025). These findings suggest that temperature trends are potentially underestimated by climate models, if these atmospheric circulation changes are due to anthropogenic forcing instead of internal variability. Thus, understanding which atmospheric dynamics associated with heat extremes are becoming more frequent over time is the first step in for understanding the mismatch between observations and projections. While individual studies have documented how certain atmospheric dynamics contribute to heat extremes over western Europe, a systematic analysis of changing dynamics of heat extremes is to our knowledge still lacking.

To investigate changes in western European heatwave dynamics, we apply the deep learning (DL) framework introduced by Happé et al. (2024) to first identify dynamically different type of heatwaves, and then study the long term changes in the latent space and their implications on the identified circulation types. The DL framework first reduces the dimensionality of spatio-temporal heatwaves using a Variational Autoencoder (VAE), before clustering the heatwaves based on the extracted features, and is trained on a large ensemble climate model dataset (Happé et al., 2024). In this research, we apply that DL framework to the ERA5 reanalyzed observational dataset (Hersbach et al. 2020), to be able to study the historical changes in dynamics. We first investigate whether the VAE model generalizes satisfactorily to the ERA5 dataset. Secondly, we examine whether the four emerging circulation types are the same across the climate model and the reanalysis. Thirdly, we examine trends in the frequency of occurrence of the different types of heatwaves. More crucially, we also investigate changes in the circulation of the heatwave types, through an analysis of the latent space of the VAE model. In recent years, DL models have become increasingly more powerful, from prediction of extreme events to foundation models of the climate (Camps-Valls et al., 2025). Yet, utilizing the full potential of such DL models to provide new insights into extreme weather events remains challenging (Camps-Valls et al., 2025). Therefore, this paper tailors and uses several interpretability methods to provide insight into the heatwave dynamics as learned by our DL models, as well as the trends associated with the different heatwave types.

We apply several interpretability methods to the DL framework, to provide insight into the clustering of the dynamically different types of heatwaves and into the changes in the latent space. While many different explainable AI (XAI) methods are available in the field of weather and climate, these are mostly used for classification and prediction tasks, and different XAI approaches can sometimes give conflicting results (Bommer et al. 2024). One example of an application of XAI on a heatwave classification task is by Marina et al. (2025), who use SHAP values on their Variational Autoencoder (VAE) model to highlight areas in their spatial input fields important for classifying heatwaves. Another way of understanding the latent space of a VAE, without specific training for a classification task, is by perturbing the input of some nodes before reconstruction. This has been shown to be useful in, e.g., the medical field (Kuznetsov et al. 2021) but also in the weather and climate community to understand circulation behind typhoons better (Hsieh & Wu, 2024). We build upon these ideas to understand which nodes in the latent space are most important for individual heatwave clusters and what impacts those nodes have on the circulation features of the heatwaves. We furthermore apply trend analysis in the latent space, and investigate the effect of shifts in the latent space on heatwave circulation.

## 2. Methods

### 2.1 Data

We use daily data from the ERA5 reanalysis dataset from the period 1940 to 2023 (Hersbach et. al, 2020). The atmospheric variables are mean sea level pressure (MSLP) and stream function at 250 hPa (STREAM250), as calculated from the u and v wind fields using Climate Data Operators (Schulzweida, 2023). We furthermore use temperature at 2 m above the surface (T2M). In this study, we focus on summer heatwaves (July-August) in the western European region (35°N–55°N, 10°W–15°E), for which we analyse the accompanying atmospheric circulation over the extended north-Atlantic sector (30°N–75°N, 75°W–60°E). The data is regridded from a 0.25 degrees resolution to ~0.7°, the grid of the EC-Earth climate model dataset used in Happé et al. (2024), LENTIS-KNMI (Muntjewerf et al., 2023), since it is essential that the dataset has the same dimensions, to use the existing deep learning model.

## 2.1 Preprocessing

### 2.2.1 Dynamical Decomposition.

We select heatwaves based on a single threshold applied over the entire time series of T2M; because of this, we need to account for the trend in T2M. Since we are specifically interested in the changing dynamics of heatwaves over time, we want to only remove the thermodynamic component from our T2M time series. In order to do so, we use the ridge regression dynamical decomposition method from Pfleiderer et al. (2025) (Figure A1). We train a ridge regression (Eq. 1) to predict T2M at a given grid point, using the stream function at 500 hPa ($\phi$) surrounding the grid point (as a proxy for the circulation) and the 5-year smoothened global mean surface temperature (GMST) (as a proxy of global warming):

$$1)\ T = \Upsilon_0 + \Upsilon_1\ GMST + \sum_{j}^{grid-cells} (\Upsilon_j \phi_j) + \epsilon$$

The streamfunction time-series of neighboring grid-cells are highly correlated and there would be a high risk of overfitting when applying a simple multiple linear regression. The ridge regression mitigates overfitting by introducing a penalty factor to the residual sum of squares for the minimization. We apply this penalty only to the streamfunction time-series (and not to the GMST coefficient and the intercept). The alpha is set to 10 for all gridpoints, to avoid overfitting. We then calculate the grid point T2M as predicted by GMST only, and remove this from the original T2M timeseries. We are then left with a T2M timeseries where the thermodynamic-driven trend is removed. Any remaining trend is then due to dynamic changes and/or natural variability. The new time series is called T2M$_{minus\ thermo}$.

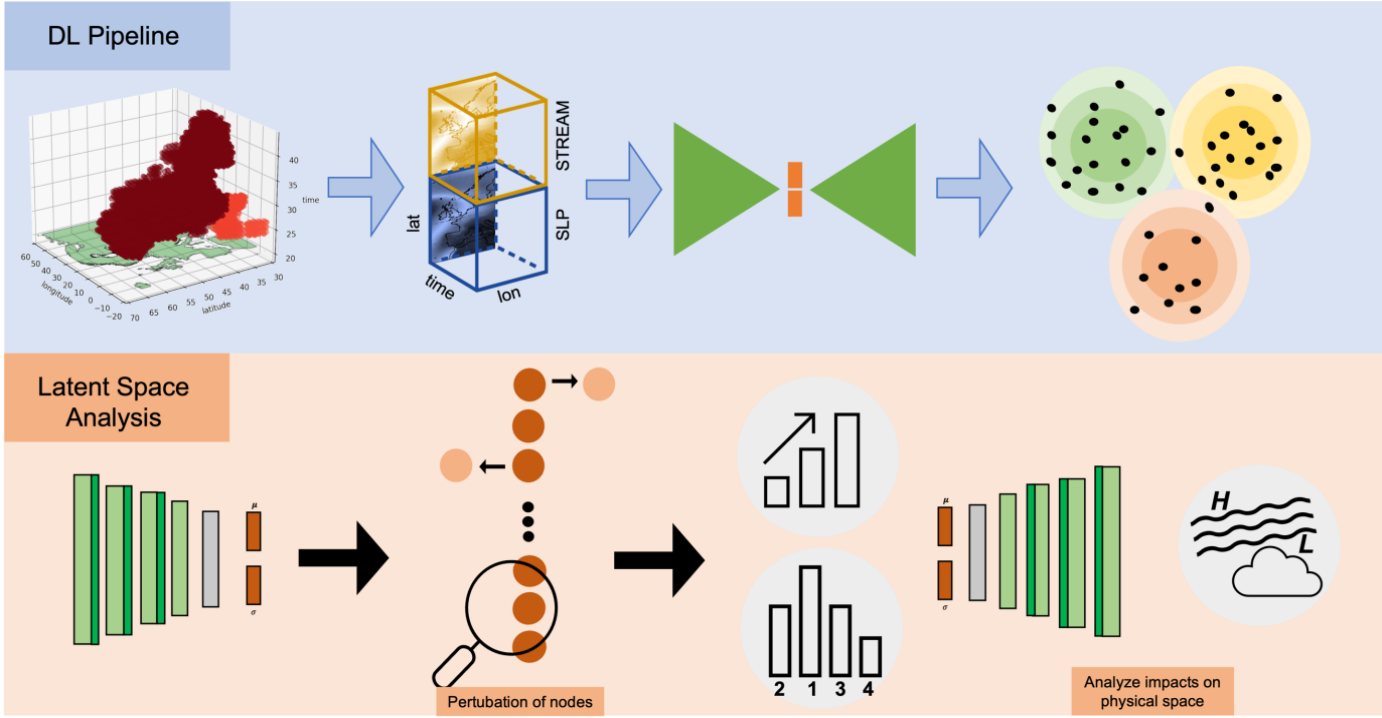

**Figure 1. Methodological Overview.** The top panel shows the Deep Learning (DL) Pipeline, following Happe et al., (2024) (section 2.3), which includes the heatwave selection, VAE dimensionality reduction, and the clustering of the heatwave samples. The bottom panel depicts the Latent Space Analysis (section 2.4), where samples are taken through the encoder part of the VAE, after which we investigate feature importance, the physical features corresponding to the nodes, and trends in the latent space.

### 2.2.1 Heatwave selection.

We select heatwaves using the GDBSCAN algorithm (Sander et al., 1998), as detailed in Happé et al. (2024) (Fig. 1). In this method, grid points that exceed their 90th percentile in T2M$_{minus\ thermo}$ time series are clustered together in space and time. A heatwave is a cluster of at least 21 grid points, allowing us to capture localized extreme heatwaves as well as large scale heatwaves. The 90th percentile is calculated specific to the grid point and to the calendar day, using the 15 calendar days surrounding the day and across all years, to account for seasonality. We obtain 739 heatwaves in ERA5 in the T2M$_{\text{minus thermo}}$ timeseries. For each heatwave we define a "heatwave circulation sample", by selecting 5-day windows from day one of the heatwave, and selecting the MSLP and STREAM250 variables over the North Atlantic region. Thus, each heatwave is associated with information about its associated circulation with the dimensions of 192 (longitude), by 64 (latitude), by 5 (time), by 2 (variables). The MSLP and STREAM250 variables are standardized using a multi-year 15-day rolling grid-point standard deviation and mean, and are cut at -3 and 3 standard deviations corresponding to 99.7% of the values according to the three-sigma rule, following Happé et al. (2024).

*2.3 Deep Learning for Heatwave classification*

*2.3.1 Generalizability of the DL Framework.* To reduce the dimensionality of our heatwave samples, we use the trained 3D VAE with 128 latent dimensions from Happé et al. (2024). This VAE was trained on heatwave circulation samples – defined as in Section 2.2 – from the LENTIS large ensemble of stationary climate data (Happé et al. 2024). We first need to assess whether the VAE with the weights as trained on LENTIS heatwave samples can accurately reconstruct the ERA5 heatwaves samples. To do so we use the R2 score of determination, to be able to compare to Happé et al. (2024) and others.

*2.3.2 Assessment of Heatwave types in ERA5.* We cluster the heatwave samples, as represented by the latent dimensions of the VAE, using a Gaussian Mixture Model (GMM). The GMM assigns to each sample probabilities of belonging to each cluster, instead of assigning clusters deterministically. We then select the n heatwave samples closest to the centroids of each cluster, to analyze the different dynamics of each cluster, defining these as central heatwaves. We also investigate the associated surface temperature imprints of each of the central heatwaves, with and without the thermodynamic trend removed.

*2.3.3 Assessment of Heatwave types in ERA5.* To assess whether heatwave types are changing over time, we calculate the trend in each heatwave cluster over the entire time series, from 1940-2023. To do so, we use a linear regression fit over the count of heatwaves per year, for each of the four clusters. We also calculate the trend for the last few decades, from 1979 to 2023. This interval corresponds to the satellite era, when climate reanalysis is deemed more accurate, and arguably to the clear emergence of the global warming signal in observed temperatures. We assess statistical significance of the trends using p-values of 0.05.

*2.4 Understanding trends in heatwaves*

The latent space of the VAE consists of 128 nodes, hence each heatwave sample can be reduced to 128 values, corresponding to a reduction in dimensionality by a factor ~1000. As we will see, this drastic reduction in dimensionality still reconstructs all individual heatwaves well. We analyse the latent space using four interpretability methods to understand the heatwave clusters and changes in the latent space (Fig. 1).

First of all, we identify the most important nodes for the cluster allocations. We do this by calculating, for each of the 128 nodes, the pairwise differences between their average values for the four clusters. We then define the *n* most important nodes to differentiate between clusters X and Y, i.e. those with the largest difference in average node-value between cluster X and Y. By changing the values of those *n* nodes for each heatwave sample in cluster Y to the value of the central heatwaves in cluster X, we try to change the cluster predictions. This process is started from the node with the largest difference and progressively repeated for nodes with smaller differences, to determine how many and which nodes are most important for the cluster allocation.

To then understand which circulation features the most important nodes represent, we change the values of these nodes for the heatwave samples in cluster X to the mean values of cluster Y, to eliminating the effect of these nodes. By plotting the difference between the newly reconstructed heatwave samples and the original ones, we can identify the physical features that belong to the most important nodes. We repeat this analysis for each of the four different clusters – i.e. to determine the most important nodes for cluster 1, we repeat this pairwise analysis with X=1 and Y=2, Y=3, and Y=4.

Second, we analyse the trends in the latent space by assessing the linear trend of each node individually. We calculate the linear trend with a linear regression model, using the years of the heatwave data from 1940–2023, and using a p-value threshold of 0.05 for significance using a two sided t-test, with p-values adjusted for the large amount of tests conducted, calculated with the false discovery rate method of Benjamini-Hochberg. To understand what circulation features these changes in the latent space represent, we add a 100-year linear increment to all heatwave samples, for those nodes that have statistically significant trends. For comparison, we also repeat this analysis using the trends of all nodes, regardless of their statistical significance. We then plot the difference between the altered heatwave samples and their original values, to assess the circulation features that belong to the trends in the latent space.

Third, we repeat the second approach, the trend analysis of the nodes in the latent space, but apply it for each cluster separately. This way, we are able to assess whether different types of heatwaves (corresponding to the clusters) have different trends.

Lastly, we use the 100-year linear increment of each individual cluster, to 'boost' just the central heatwave samples. Doing so, we visualize how the circulation of each type of heatwaves could potentially look like in the future, if the linear trends continue.

**3. Results**

*3.1 Dynamically different type of heatwaves in ERA5*

*3.1.1 Generalizability of the DL Framework.* Using the weights from the trained model, we obtain a competitive reconstruction error for ERA5 heatwave samples, compared to the LENTIS validation set, with an average $R^2$ score over all heatwave samples of 0.889 and 0.900, respectively for ERA5 and LENTIS (Fig. 2A). However, the left tail is slightly thicker for the ERA5 heatwaves, indicating a few heatwaves in this dataset are reconstructed less well. Still, while the amount of compression is very high and the dataset is different than the one used during training, we observe that the main atmospheric circulation features are reconstructed well by the VAE (Fig. 2B). This indicates that the extracted features in the latent space represent meaningful features of the heatwave circulation. The high reconstruction scores and features (Fig. 2) give us confidence that the VAE can directly be applied to ERA5 without fine tuning.

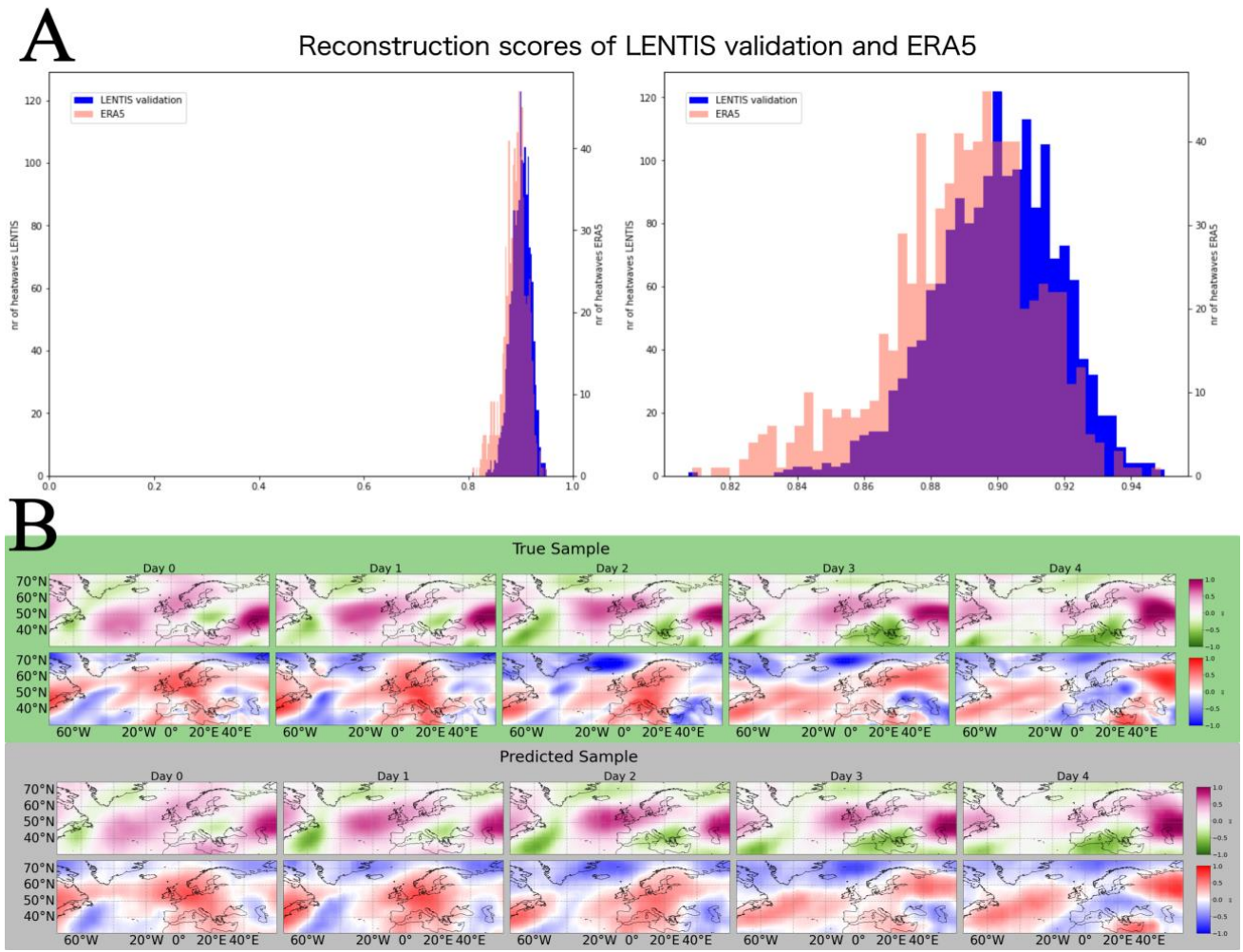

**Figure 2. Reconstruction of ERA5 heatwaves using VAE. A)** Distribution of reconstruction error (R2) of all heatwave samples in ERA5 and those of the LENTIS validation set, zoomed in on the right panel, and **B)** an example of a reconstructed heatwave sample (bottom) versus the original (top), for the 5 days and 2 variables (STREAM250 on top and MSLP on the bottom row).

*3.1.2 Heatwave clusters.* The four different clusters, following the Gaussian Mixture Model clustering, are similar to those found with the LENTIS data in Happé et al. (2024). Here we show the composite of the 5 heatwaves closest to each centroid (Fig. 3), which we define as 'central heatwaves', representing each cluster. The temperature anomalies corresponding to these central heatwaves are shown in Figure A2. The first cluster is the Atlantic Plume, which consists of a low-pressure system in the North Atlantic Ocean off the coast of Portugal and an extended ridge structure over southern Europe. The temperature imprint shows a very high temperature anomaly over Portugal and Spain, which extends towards the north after a few days. Indeed, this type of circulation is known to cause advection from the south bringing warm air, at times even reaching London (Sousa et al., 2019; Yiou et al., 2020). The second cluster is the UK high, consisting of a persistent high-pressure system located over the UK and central Europe, evident both in the upper and surface level atmospheric circulation. Surprisingly the temperature anomaly seems to be less persistent and moves from the UK eastwards, whereas this type of circulation typically is associated with persistent temperature anomalies over the UK (Sousa et al. 2021; Holmberg et al., 2023). The third cluster is the Atlantic Low, showing a high-pressure system over the wider Scandinavian region and a low-pressure system in the Atlantic, evident in both the upper and lower atmosphere. The temperature anomalies of this heatwave type show a persistent high anomaly in Portugal, which could also be a result of advection (Yiou et al., 2020; D'Andrea et al., 2024). Lastly, the fourth cluster is the Atlantic High, which shows a general higher sea level pressure over the entire European and Atlantic region. This circulation type is accompanied by a general positive temperature anomaly across the entire southern part of western Europe.

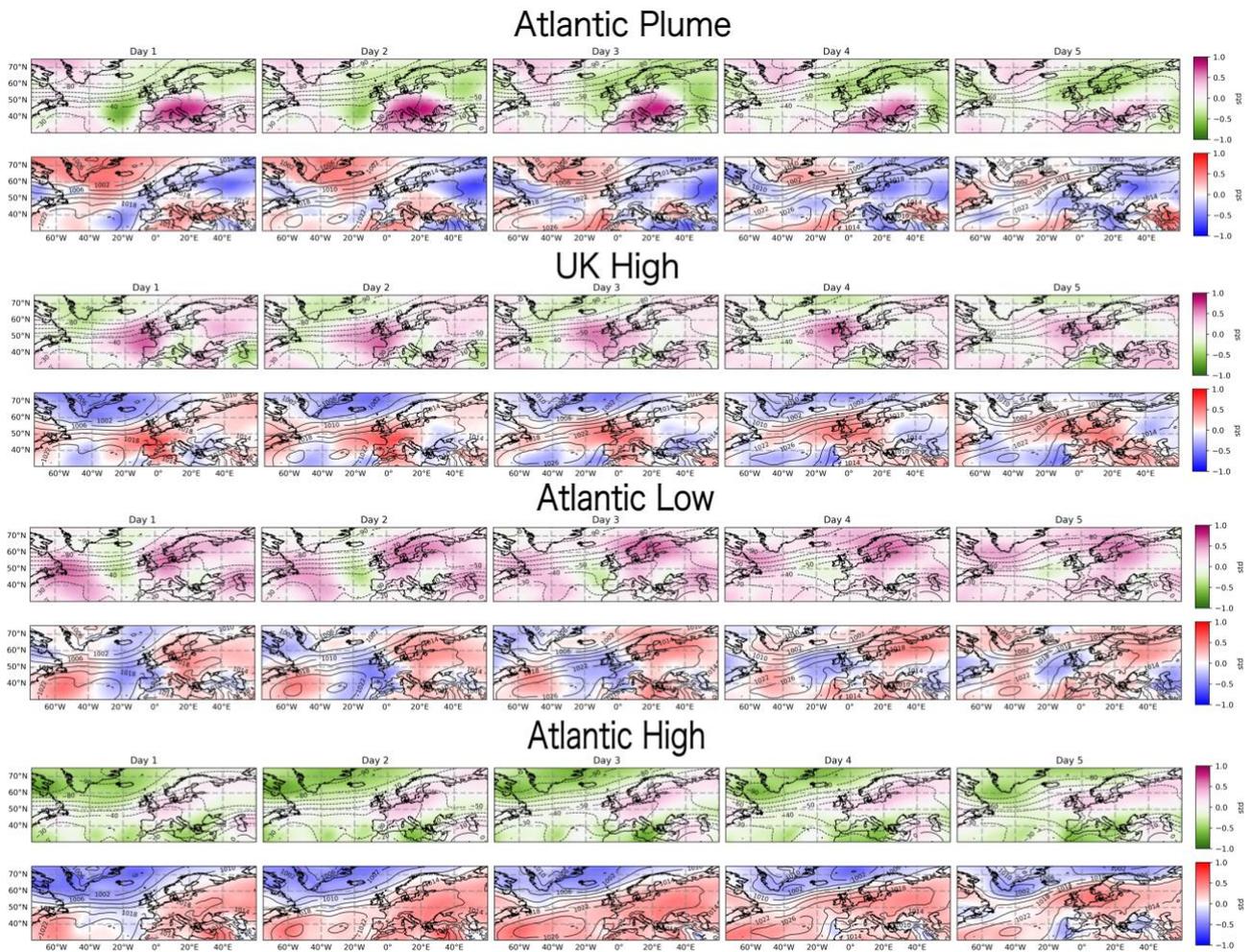

**Figure 3. The four central heatwaves of ERA5,** here plotted as the mean of the five closest heatwaves to each centroid. The colors depict the anomalies with respect to the 15-day climatology, in standard deviations, and the contours show the original values. For streamfunction at 250hPa the values are in $10^6$ m$^2$/s and for mean sea level pressure the values are in hPa.

*3.1.3 Trends in heatwave types.* First, we count the amount of heatwave samples that belong to each of the four clusters for each year. We then assess whether there are trends in these amounts, by applying a linear regression for different time periods (Fig. 4). We detect no statistically significant trends over the entire time period, spanning 1940-2023, even though the Atlantic Plume does show a slight increase. However, from 1979 onwards the trends become more evident, and Atlantic Plume heatwaves, have become more frequent, while the Atlantic High type have become less frequent over time. Although the Atlantic Low cluster, does not show statistically significant results, we still see a slight increase since 1979.

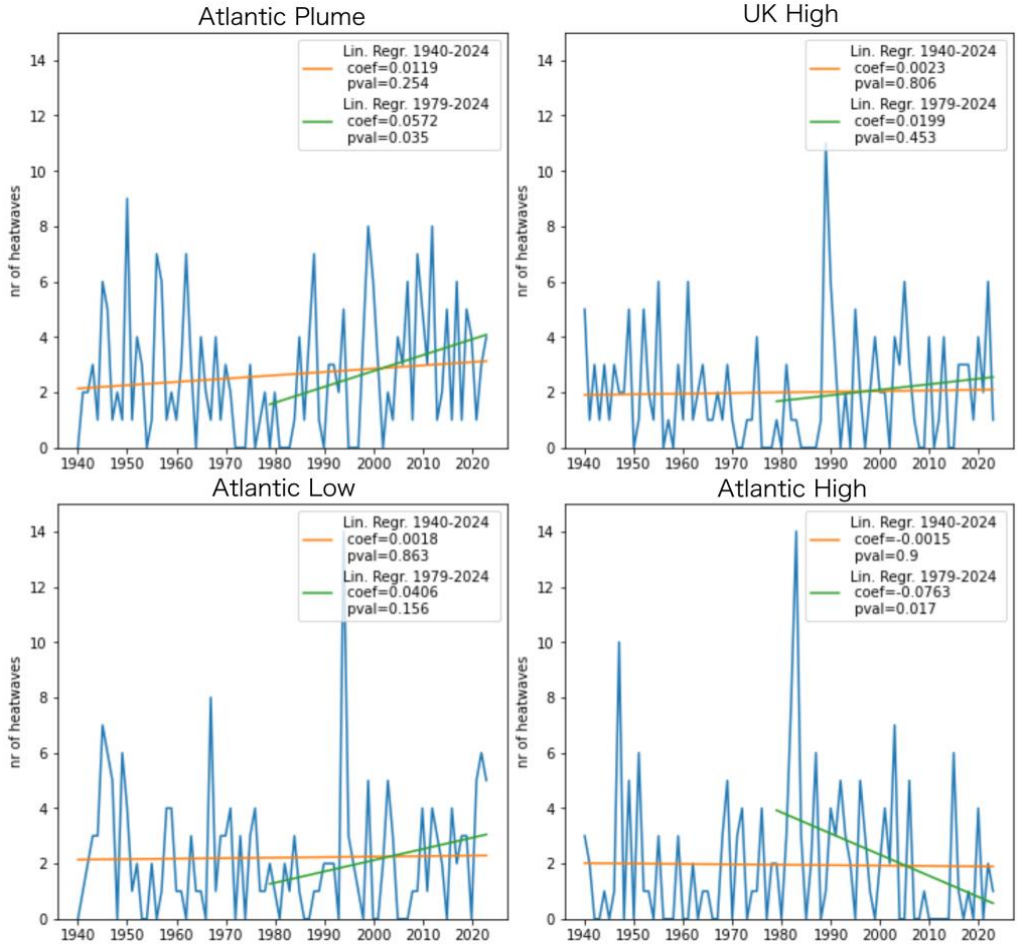

**Figure 4. Trends in different heatwave types.** In orange for 1940-2023, and in green for 1979-2023.

*3.2 Interpretation of Latent Space*

Here, we explore the link between the latent space representation of each heatwave cluster, and the physical features of these heatwaves. To understand the sensitivity of the cluster assignment to the different nodes in latent space, we incrementally change the value of the nodes of a single heatwave to the values of the central heatwaves of the target cluster, to try to convert it to that cluster (see Methods section 2.4). Figure 5A shows that a 100% conversion rate is reached for the heatwaves in a single cluster when changing all 128 nodes, as expected. One can also see that by changing only a few nodes, a large fraction of heatwave samples can already be converted to cluster 1; e.g., by changing the value of 10 nodes, ~30% of heatwave samples are converted. For clusters 3, and 2 and 4 the conversion curve is less steep, indicating more nodes are needed to distinguish the clusters. These results indicate that not all of the 128 nodes are equally important for the clustering task. To investigate which circulation features belong to the most important nodes, we take the 20 most important nodes and change their values to those of a the central heatwaves of a different cluster. By reconstructing the heatwaves with those changed values, we can determine which circulation features have changed. Figure 5B shows that indeed the most important features align very well with the key elements that can be found in the central heatwaves. This confirms that the clustering algorithm differentiates heatwaves based on these key circulation features. For example, Cluster 3 - The Atlantic Low shows a negative anomaly in SLP and STREAM250 and a positive anomaly above Scandinavia (Fig. 5B), features that can be seen in the central heatwaves as well (Fig. 3B). Similarly, Cluster 2 – The UK High shows a feature consisting of a positive anomaly over the UK, also evident in Figure 3B.

We then investigate the trends in the latent space, by analysing trends of each node. In this way, we are able to determine how the atmospheric circulation of heatwaves are changing (Methods section 2.4). For those nodes that show statistically significant trends (in red, Fig. 6A), we calculate a 100-year increment value and add this to each heatwave sample. We then plot the reconstruction differences between the changed samples and the original ones, to assess which circulation correspond to the latent space changes (Fig 6B). Here we find that Greenland experiences a positive trend in both streamfunction and sea level pressure, while the North Atlantic Ocean experiences a negative trend in sea level pressure – both consistent with observed trends. We furthermore see a plume structure in the streamfunction, associated with lower trends over Scandinavia. Lastly, Figure 6B shows an increase in sea level pressure over Russia. These results are similar, if not more pronounced, when we take the 100-year increment of all 128 nodes in the latent space, irrespective of statistical significant trends (Fig. A3).

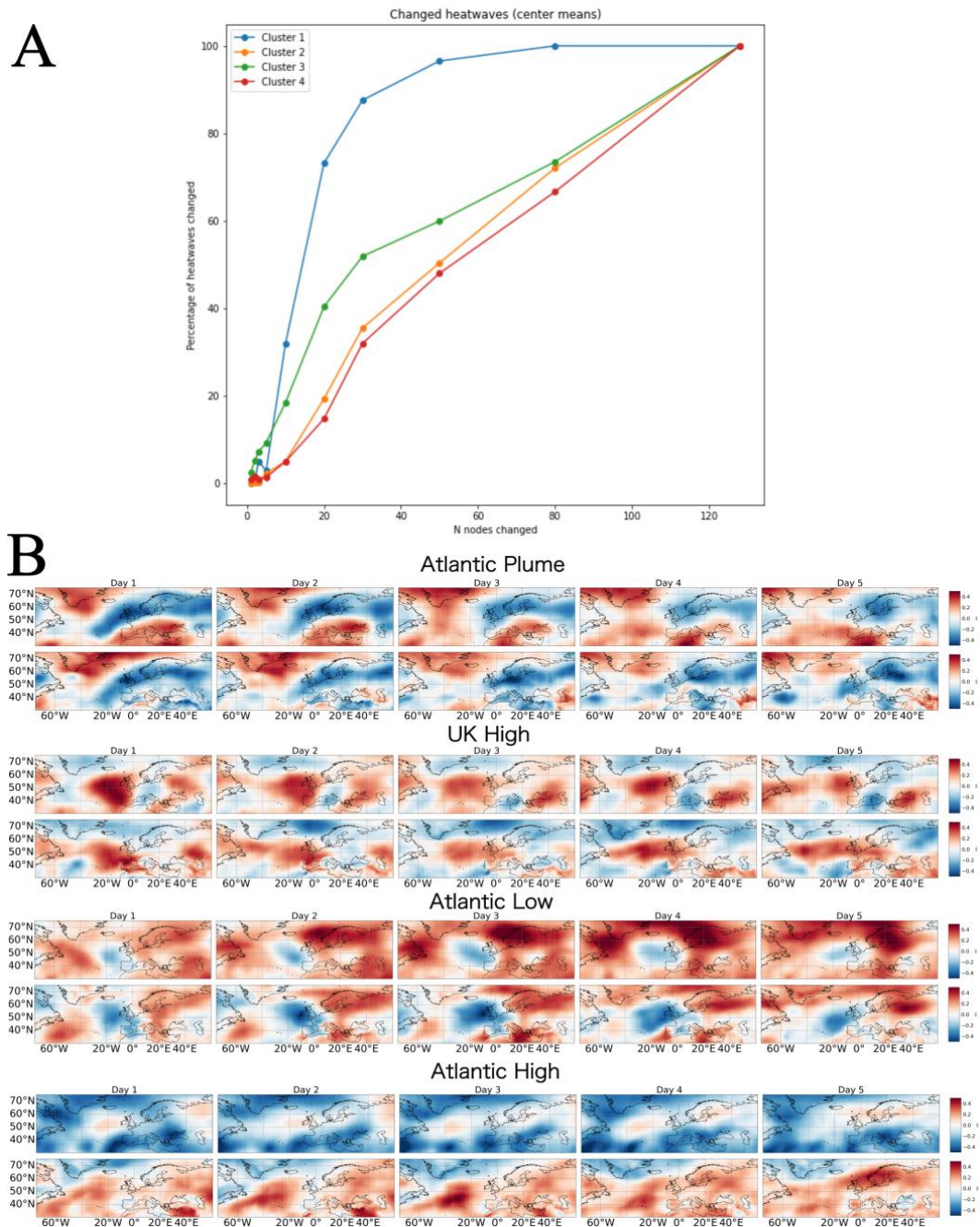

**Figure 5. Feature importance (A) and identification (B).** A) number of nodes needed to convert y-percentage of heatwaves to cluster 1, 2, 3 or 4 respectively, and B) physical features associated with the 20 most distinctive nodes, shown here for the central heatwaves.

We repeat the latent space trend analysis for each cluster separately, determining which nodes are changing significantly and which circulation features belong to those nodes (Fig 6C). Here we see that the changes are not uniform over the dynamically different types of heatwaves, the changes are furthremore much stronger for the individual types than for all combined, which could be the result of averaging out trends when combining the different heatwave types. The Atlantic Plume cluster shows a decrease in MSLP in the southern part of the North Atlantic, and the increase in southern plume visible in the overall trend in the upper atmosphere is less evident in those heatwaves belonging to this cluster. On the other hand, the positive streamfunction changes are more evident in the UK High cluster, with a slight shift of the high pressure eastward. The Atlantic Low shows a decrease in both MSLP and streamfunction in the North Atlantic, while the MSLP over Scandinavia increases. Lastly, the fourth cluster – the Atlantic High shows an increased wave-like structure, following a high - low - high pattern.

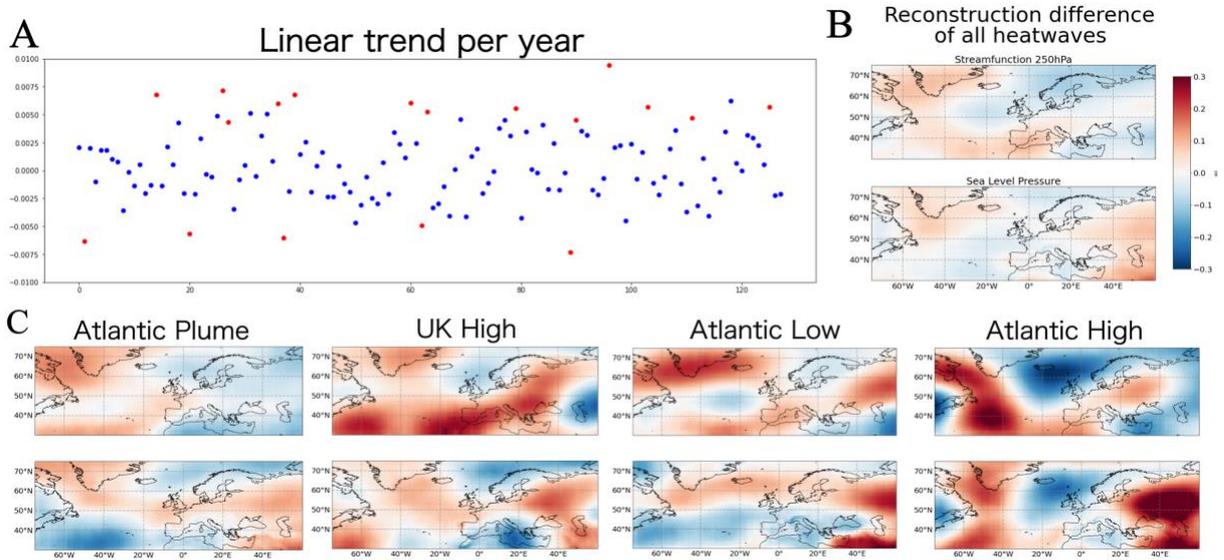

**Figure 6. Latent space trends and its physical features. A)** trends per node for all heatwaves combined, red dots indicate statistically significant linear trends with adjusted p-values below 0.05, **B)** physical features associated with the trends (5 day mean fields), in standard deviation. The top shows the streamfunction at 250hPa and the bottom shows the MSLP, and **C)** the same as B but the trends are calculated for each cluster separately.

Then, we take the central heatwaves and 'boost' each of them with a 100y increment based on their own cluster changes, to see what a future heatwave would look like (for the assumption that the trend continues). Figure 7 shows how some patterns become amplified, like the blocking structures in the Atlantic Plume and the UK High. It is also possible to see new features emerging, for example in the Atlantic High cluster the streamfunction shows amplification of wave-like structures over the North Atlantic. Mostly, we observe that the circulation changes accompanying the trends (as shown in Fig. 6C) are added to the original heatwave circulation (as shown in Fig. 3B.

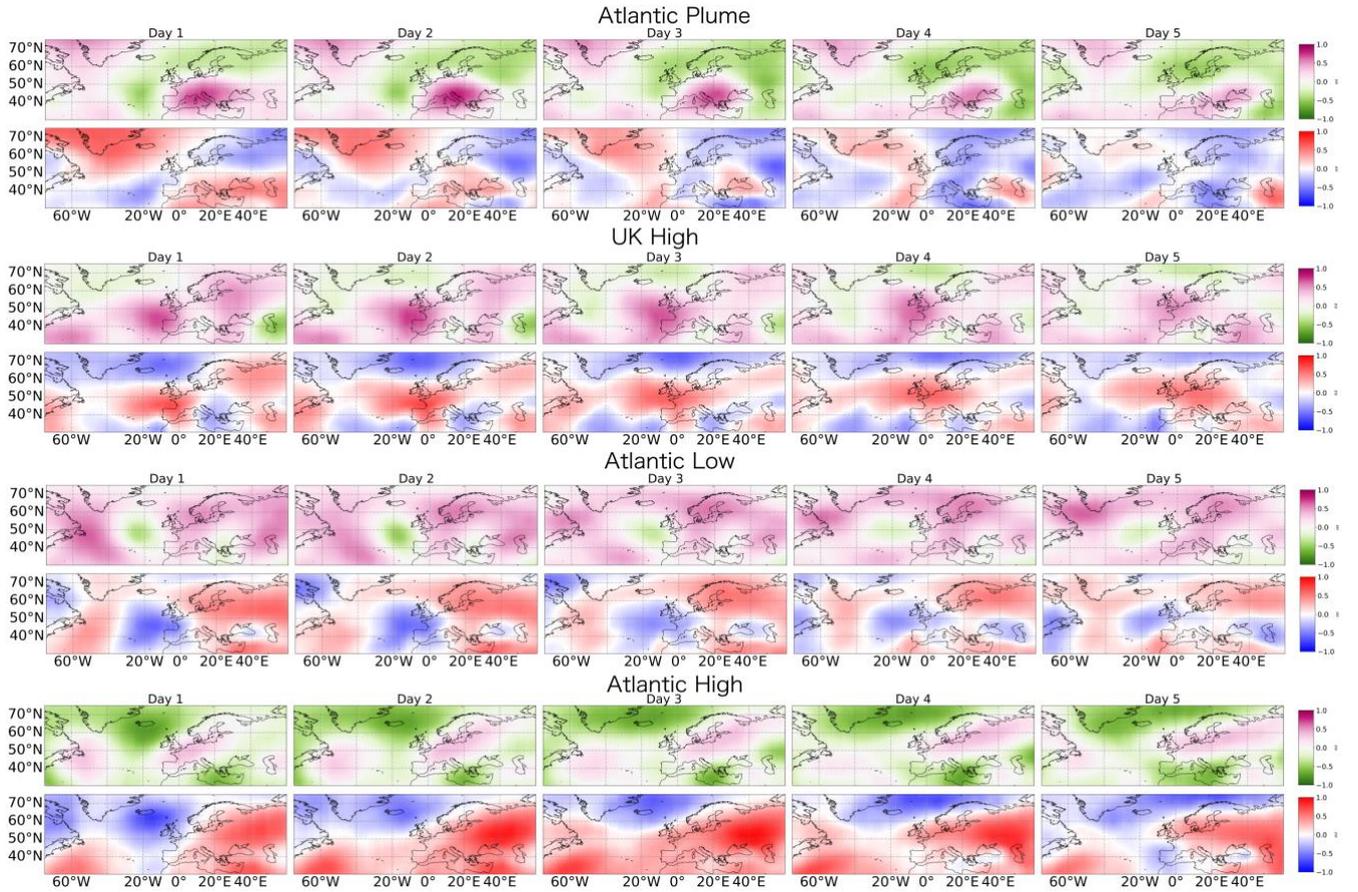

**Figure 7. Boosted central heatwaves,** with the latent space trends as shown in 6C.

## 4. Discussion

*4.1 Trends in atmospheric dynamics of European heatwaves*

By clustering the latent representations of the ERA5 heatwave samples, we find the same types of heatwaves as Happé et al. (2024) did using large ensemble climate model data. Those four types of dynamically distinct heatwave types – Atlantic Plume, UK High, Atlantic Low, and Atlantic High — and their temperature imprints are consistent with previous literature (Stefanon et al., 2012; Sousa et al., 2019; Yiou et al., 2020; Sousa et al., 2021; Holmberg et al., 2023; Rouges et al., 2023). Through different mechanisms, each type is able to bring warm temperature anomalies to western Europe, either through advection (e.g. Atlantic Low/Atlantic Plume) or through blocking systems (UK High/ Atlantic High). Next, we find statistically significant changes in the frequency of the heatwave types from 1979 onwards. The Atlantic Plume cluster is increasing over time, which consists of heatwaves with a low MSLP anomaly in the North Atlantic west of Portugal in combination with a positive anomaly in the streamfunction in the form of a plume (i.e. a 'Saharan plume' circulation advecting warm air northwards towards western Europe). These findings are in line with previous studies, which show both an increase in frequency of such deep lows (D'Andrea et al., 2024) and in plume like advection (Sousa et al., 2019; Vautard et al., 2023). In contrast, we find that the Atlantic High cluster is decreasing in frequency, aligning with the increasing Atlantic Low type; this could potentially be explained by a shift in atmsopheric wave acitvity in the northern hemispheric midlatitudes (Happé et al., 2025). While trends are found from 1979 onwards, we did not find significant trends for the full period (1940-2023). It needs to be noted that before 1979 the reanalyzed ERA5 dataset is of less quality, since it was before the remote sensing era. Another possible explanation can be found in aerosol emissions, which peaked over Europe around 1979. The period 1979-2023 represents a period of strong decline in aerosol emissions and the identified changes in heatwave-cluster frequencies could very well be related to regional cganging climate processes due to the decline in aerosol concentrations (Chemke & Coumou, 2024; Kang et al., 2024; Schumacher et al., 2024).

With our latent space analysis, we find that the overall circulation changes of all heatwaves – Greenland High, Saharan Plume, and Atlantic Low – show very similar patterns to previous research investigating the relation between changes in dynamics and temperature (Singh et al., 2023; Pfleiderer et al., 2025; Happé et al., 2025). This supports the validity of analysing trends in the latent space to explore changes in dynamics. Each dynamical heatwave type shows different changes over time, each showing more amplified trends than when considering the trend of all heatwaves. While some similarities can be seen between the trends of individual types and that of all heatwaves, our results emphasize the the need to understand how each type of dynamical heatwave is changin individually. This analysis could be expanded further in future research, to more precisely understand which trends in which nodes are responsible for the observed circulation changes. As each atmospheric circulation type has different causal drivers and mechanisms leading to warm temperature anomalies, if and how their circulation features may change over time can very well be different too. Understanding the individual circulation changes is crucial to understanding the potential drivers of these changes, and whether this is due to internal variability or external forcing. Especially in the context of the influence of circulation changes on temperature trends in Europe and beyond, as the drivers behind these changes still remain unclear (Shaw et al., 2024). Research indicates possible connections to aerosols (Chemke & Coumou, 2024; Kang et al., 2024; Schumacher et al., 2024), SST variability (Happé et al., 2025), and soil moisture feedbacks (Beobide-Arsuaga et al., 2025; Tian et al., 2025), but more research is still needed.

*4.2 Interpretable deep learning*

Happé et al. (2024) used 1600 years of stationary data to train a 3D VAE model to extract heatwave features, and use data augmentation techniques to avoid overfitting. Here, we have shown that this VAE model is able to generalize to ERA5 heatwaves well, with competitive $R^2$ scores compared to the LENTIS validation set from Happé et al. (2024). We hypothesize that the data augmentation techniques play a big role in creating a generalizable model. Our reconstruction scores are much higher than for training on just ERA5 data (e.g., Paçal et al., 2025). Our successful generalization from climate models to reanalysis datasets, using data augmentation, potentially suggests that such techniques can be promising to create DL models that are generalizable to similar but unseen events and datasets. Generalizability is even more important when considering future climate, since we will be dealing with never-seen, out-of-sample data. This is higlighted in forecasting problems as well, as numerical weather prediction models still outperform the current DL weather models when it comes to forecasting extreme events (Zhang et al., 2025). One of the potential causes to this is that DL models struggle with generalization, in this case forecasting unprecedented events outside their training data (Zhang et al., 2025), which is where data augmentation could potentially improve these models. Lastly, analysing differences in the latent space between different datasets can allow us to better understand differences between past and future (as we show in this study), but potentially also between climate models and reanalysis data in future research.

We show that deep learning models can accurately capture the most important circulation features of heatwaves in their latent space, where these features can be used to cluster the heatwaves into meaningful clusters. Moreover, we show that the information in the latent space can provide insights into spatio-temporal changes of these heatwave dynamics and the most important circulation features that characterizes each type of heatwave cluster. While deep learning methods have shown serious improvement over the last few years in climate sciences, most models still focus on improving skill in forecasting weather. Those studies that use DL to better understand the climate system mostly use existing XAI techniques. However, these techniques can sometimes show inconsistent results and thus need careful consideration when applying these for the understanding of climate questions (Bommer et al., 2024). For example, SHAP values interpretation is always in relation to the refrerence data, and a postive SHAP thus does not mean that an increase in that feature leads to an increased prediction. Here, we focus on understanding the latent space by devising interpretability methods that are understanable and transparent. We believe this to be a fundamental part of research on AI and weather and climate, as overcoming the "black box" nature of DL requires transparent and interpretable methods, preferably guided with expert knowledge.

Through the use of our clustering algorithm and dimensions in the latent space, we are able to identify the most important circulation features that separate the heatwave samples into different clusters. By nudging our heatwave samples towards a different cluster, through changing the nodes in the latent space, we are able to determine which nodes are the most important for cluster assignment. We find that not all nodes are equally important, and that the 20 most important nodes encapsulate the key circulation features of each cluster – indicating that indeed the clustering of the heatwaves is done based on meaningful features well represented in the latent space. This method is transparent and elegant, allowing one to exactly trace back which nodes are important for the heatwave clustering and the circulation features associated with those nodes. While XAI methods are very commonly used in classification tasks, explainable AI methods are to our knowledge much less used to understand the latent space of a (V)AE – or other DL method – and the respective clustering of the latent space. Sparse auto-encoders can also be used to create an interpretable latent space (King et al., 2025), yet their downsides are that they are generally harder to train, may need a higher number of nodes in the latent space, and are not probabilistic. Feature identification for VAEs has showed promising results in both the medical field (Kuznetsov et al. 2021) and the extreme weather community (Hsieh & Wu, 2024), and we built upon such examples and apply a more systematic approach for identifying the most important nodes and its associated physical features.

The identification of important nodes paves the way for boosting heatwave circulation in the latent space. When boosting heatwave circulation by amplifying certain nodes in the latent space, we see amplification of features as well as newly emerging features. While this shows a promising new avenue of boosting methods (Gessner et al., 2022), great care should be placed on evaluating the physical plausibility of such a method. In contrast to numerical models which are based on physical equations and parameterizations, our VAE model is not physically constrained. However, physically constrained DL models are becoming more popular, and thus a DL boosting application could be promising. Such application would be computationally much cheaper than numerical boosting methods, and could provide an alternative method to create 'dynamical' storylines for extreme events. We show here that such a method can have promising results, and encourage future research to focus on establishing a physically constrained DL boosting model.

## 5. Conclusion

We show that pre-training a deep learning VAE model on climate model data in combination with data augmentation leads to a successful generalizable VAE model, where ERA5 reanalysis data has comparable reconstruction errors as the original validation data set without the need of transfer learning. We furthermore find consistent results in our heatwave clustering analysis, with the same four types of dynamically different types of heatwaves found in ERA5 as in the large ensemble climate model dataset (Happé et al., 2024). Regression analyses indicates an statistically significant increase in the Atlantic Plume type of heatwaves, consistent with previous literature, and a decline in frequency in the Atlantic High. Moreover, through analysing changes in the latent space we are able to highlight more precisely which dynamics of the heatwave circulation are changing over time – both for all heatwaves combined as well as for each individual heatwave type. This research thus highlights the importance of understanding the changes in the different types of atmospheric circulation driving heatwaves in western Europe.

We apply straightforward and transparent interpretability methods for deep learning models, needed to understand the outcomes of dimensionality reduction and clustering models used in weather and climate research. We use feature importance in the VAE latent space in combination with our clustering model, to highlight which nodes are most important for cluster allocation, and which circulation features these nodes represent. We furthermore use the latent space to analyse changes in the associated circulation features of the heatwaves, and explore a method that leverages these changes to boost heatwave circulation dynamics into the future, by nudging specific nodes in the latent space. Our work not only shows how such methods can aid in better understanding the underlying data, in this case heatwaves, but it also shows how such deep learning methods can be used to develop new types of boosting applications for unseen extreme weather events and extrapolation into the future.

## Acknowledgements

This study was partly supported by the European Union's Horizon 2020 research and innovation program under Grant Agreements 101003469 (XAIDA project). This work used the Dutch national e-infrastructure with the support of the SURF Cooperative using Grant EINF-15260.

# Appendix

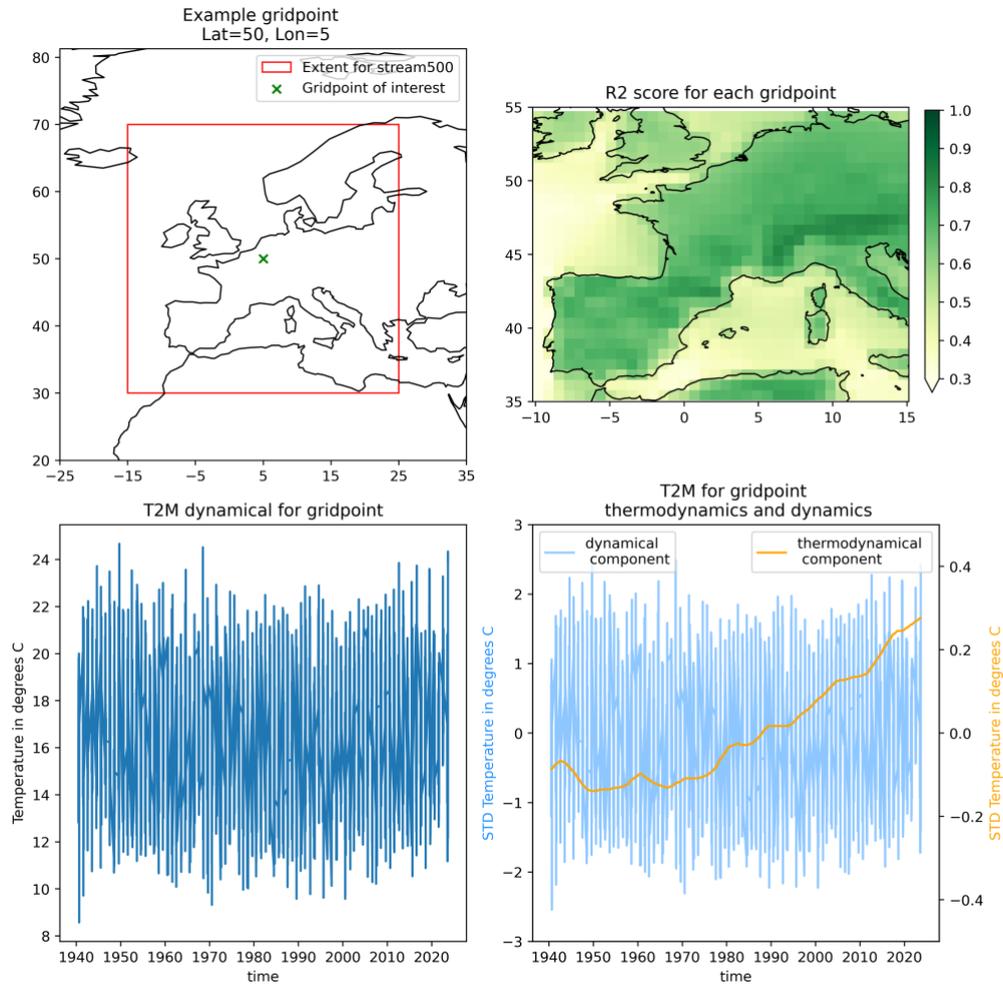

**Figure A1. Dynamical Decomposition to remove the thermodynamical component from the temperature timeseries - example for one gridpoint.** The top left shows the gridpoint of interest (in green X), and the streamfunction area around the gridpoint (red box) used for the dynamical decomposition. The top right shows the R2 coefficient of determination, indicating the proportion of variance explained by the ridge regression model, for each gridpoint in the area. The bottom row shows an example of the temperature time series of the one example gridpoint, as shown in the top left panel. The left shows the dynamical temperature as constructed using only the streamfunction and the right shows the decomposed T2M, in STD, for the two different components.

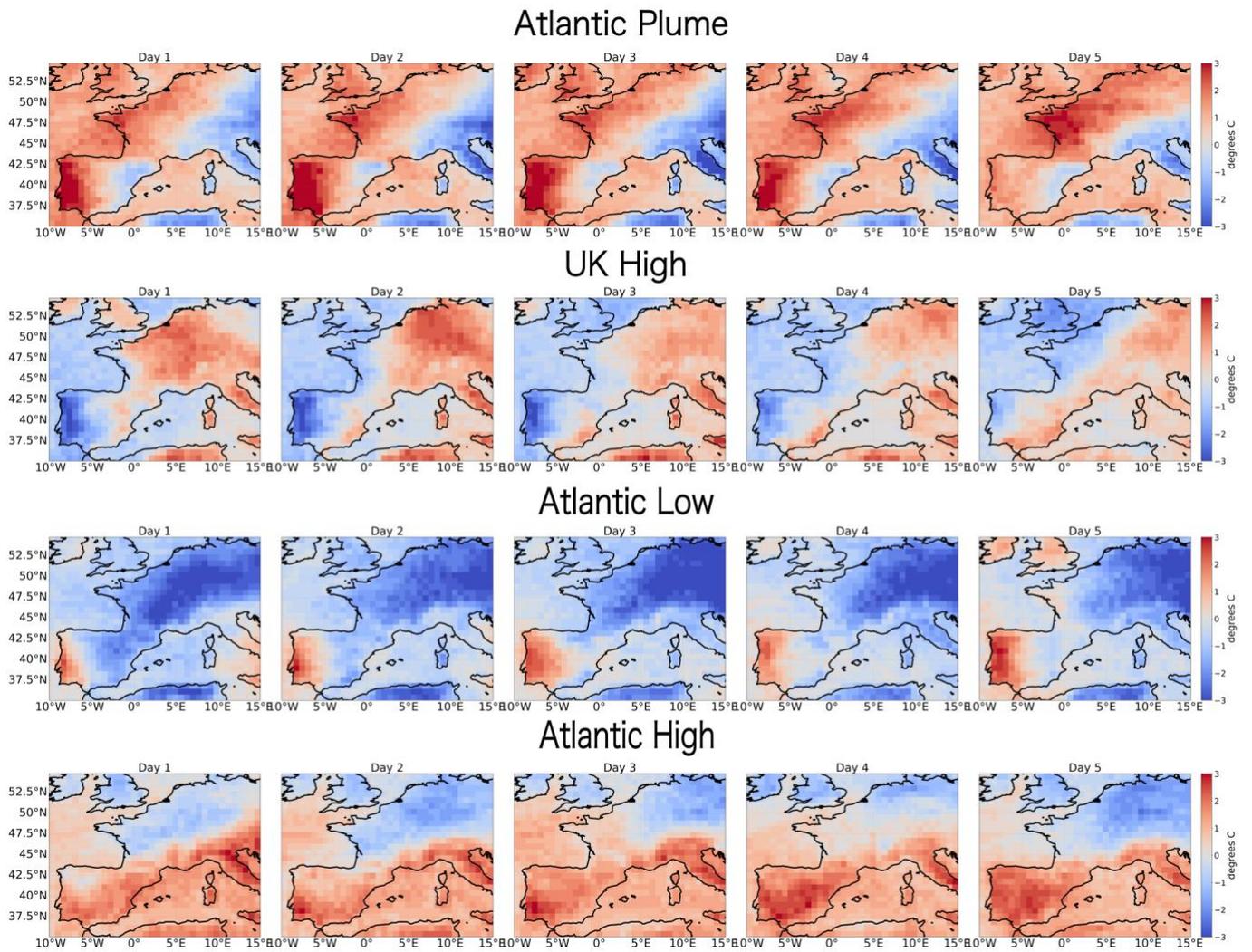

**Figure A2**. Surface temperatures (T2M) associated with the central heatwave samples of each heatwave type, as composites of the 5-closest samples to each centroid (Fig. 3).

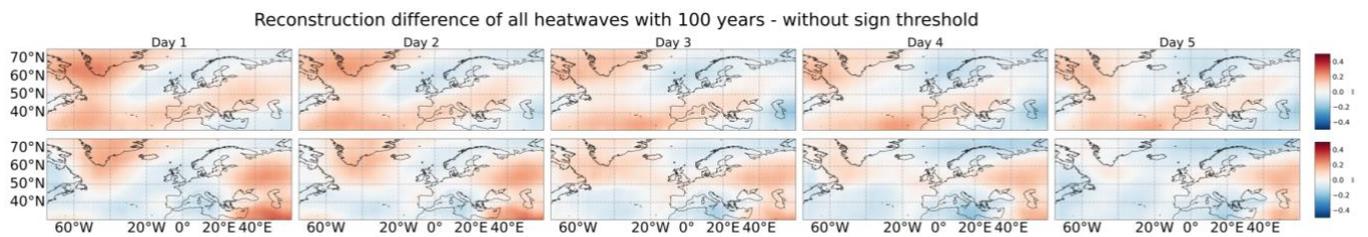

**Figure A3**. Reconstruction difference of 100y increment of the trends in the latent space, for all 128 nodes. Similar to Figure 6B, but without taking the 5 day mean.